\documentclass[twocolumn,bibnotes,showpacs,pra]{revtex4}
\usepackage{amssymb}
\usepackage{amsmath}
\usepackage{graphicx}
\usepackage{epstopdf}

\begin{document}

\title{Strongly interacting mesoscopic systems of anyons in one dimension}
\author{N.~T. Zinner}
\affiliation{Department of Physics and Astronomy, Aarhus University, DK-8000 Aarhus C, Denmark}

\begin{abstract}
Using the fractional statistical properties of so-called anyonic particles, we present
solutions of the Schr{\"o}dinger equation
for up to six strongly interacting particles in one-dimensional confinement that 
interpolate the usual bosonic and fermionic limits. These solutions are exact to 
linear order in the inverse coupling strength of the zero-range interaction of our model. 
Specifically, we consider two-component 
mixtures of anyons and use these to eludicate the mixing-demixing properties of both 
balanced and imbalanced systems. Importantly, we demonstrate that the degree of demixing 
depends sensitively on the external trap in which the particles are confined. We also 
show how one may in principle probe the statistical parameter of an anyonic system by injection
a strongly interacting impurity and doing spectral or tunneling measurements.
\end{abstract}
\pacs{03.65.Ca,67.85.Pq,71.10.Pm,05.30.Pr}

\date{\today}

\maketitle

\section{Introduction}
In the quantum world we classify physically identical particles according to 
their statistical properties and typically divide them 
into two distinct sets. The key characteristic is that upon exchange of two such 
particles the total wave function changes only by a sign which is 
positive for bosonic and negative for fermionic particles. It came as 
quite a surprise to many when Leinass and Myrheim \cite{lm1977} (see also Refs.~\onlinecite{wilczek1982,goldin1981}) 
discovered that in 
two dimensions (2D) one can accomodate exchange statistics 
that is neither bosonic nor fermionic but rather interpolates 
the usual possibilities and gives rise to so-called {\it anyonic}
particles. Systems that display effective anyonic statistics are
a topic of great current interest due to the integral role they 
enjoy in the field of quantum computation \cite{pachos2012} (see Ref.~\onlinecite{sarma2008}
for an overview of recent theoretical and experimental progress).

An early breakthrough in the understanding of anyons was achieved by 
Haldane who generalized the 2D case and introduced 
the notion of 'fractional statistics' in any dimension \cite{haldane1991}.
Anyons also play a prominent role in exploring the connection between 
statistical mechanics and random matrix theory \cite{alonso1996,jain1997}. 
In one dimension (1D), the famous Calogero-Sutherland (CS) model \cite{calogero1969,sutherland1971}
provides an example of fractional statistics and anyons \cite{poly1989}. 
The CS model can even be extended to exactly solvable many-body models exhibiting 
long-range order in 1D \cite{auberson2000}. 
Thus, 1D anyonic systems remains a research topic of 
great interest in several different fields \cite{zhu1996,kundu1999,batchelor2006,
girar2006,averin2007,feiguin2007,patu2007,trebst2008,campo2008,fidkowski2008,greiter2009,santa2008,bellazzini2009,santos2012}.
Most recently, realization of anyonic behavior in cold atomic gases have 
been proposed in both 2D \cite{paredes2001,aguado2008,jiang2008} and 1D \cite{keilmann2011} setups.
Such proposals typically require manipulation of small atom numbers. 
It is therefore encouraging that preparation of desired mesoscopic 
system sizes is becoming increasingly more precise \cite{serwane2011,wenz2013,pellegrino2014,nogrette2014}.

\begin{figure}[ht!]
\centering
\includegraphics[scale=0.33]{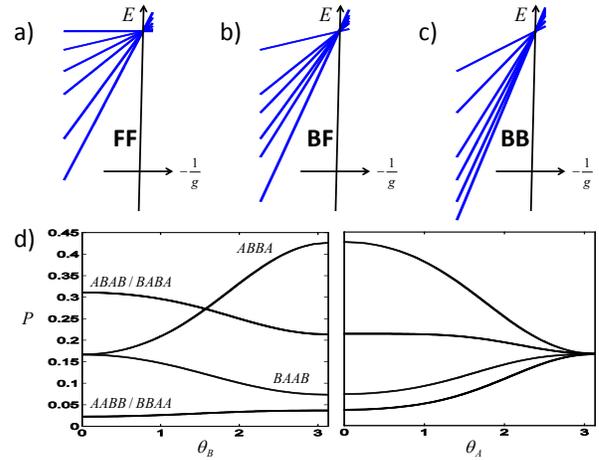}
\caption{(Color online) Four-body system of two $A$ and two $B$ particles in a hard-wall box with statistical 
parameters $\theta_A$ and $\theta_B$ respectively. The exact solution for large
short-range interaction strength, $g$, have linear spectral slopes
as shown schematically for a) Fermi-Fermi (FF), b) Bose-Fermi (BF), and c) Bose-Bose (BB) 
mixtures. 
d) The probabilities of different particle configurations for 
the ground state is shown as a function of first $\theta_B$ (keeping $\theta_A=0$) and then $\theta_A$ 
(keeping $\theta_B=\pi$). The particle configurations of each line is shown in the 
left-hand panel. The system is dominantly antiferromagnetic in the FF limit ($\theta_B=\theta_A=0$), 
then becomes demixed in the BF case ($\theta_B=\pi$ and $\theta_A=0$), 
and finally becomes completely mixed in the 
BB limit ($\theta_B=\theta_A=\pi$) where all configurations are equally likely.}
\label{slopes}
\end{figure}

In this paper we will utilize the notion of anyons to elucidate the
behavior of strongly interacting mesoscopic 1D systems. In particular, we will
describe a general framework that can deal with mixtures of 
several components of anyonic particles with the important limiting
cases being Fermi-Fermi, Bose-Fermi, and Bose-Bose systems. Using 
solutions of the Schr{\"o}dinger equation for the up to six-body systems in both box and 
harmonic confinement, we will show how statistics and trapping 
potentials are important for the tendency of two-component systems
to either mix or phase separate when the particles have strong 
short-range repulsive interactions. This mixing-demixing transition
remains a very active research area with several open questions
\cite{das2003,cazalilla2003,adilet2006,fang2011,wang2012}. 
The results we present quantify exactly what one should 
understand by demixing at the level of particle ordering in the exact
wave functions that take the full trap geometry into account and thus
go beyond any local density approximation based on Bethe ansatz, mean-field, 
or Luttinger liquid theory.

The solutions of the Schr{\"o}dinger equation are obtained based on a recently 
developed functional approach 
to systems with strong zero-range interactions parametrized by 
a coupling strength $g$. The solutions we present are exact to 
linear order in $1/g$.
It does not rely on Bose-Fermi \cite{girardeau1960,cheon1999,girar2004,girar2010}
or Anyon-Fermi mappings \cite{girar2006} as these techniques are not 
capable of solving general multi-component systems \cite{vol2013}. 
In contrast, our approach yields energies and 
wave functions that are adiabatically connected to the eigenstates
for large by finite interaction strengths. In Fig.~\ref{slopes}
we show an example with four particles in a hard-wall box 
(open boundary conditions) for 
different particle statistics (to be defined below).
In order to reach the strongly interacting ('hard core') regime 
one typically tunes the interaction strength from weak to 
strong in experimental setups.
It is therefore essential to provide theoretical predictions that 
take the preparation into account. This is not possible if one starts
from the strictly impenetrable 
so-called Tonks-Girardeau \cite{tonks1936,girardeau1960} regime
where all manipulations are done assuming an infinite short-range
repulsion. Our framework naturally provides suggestions for probing 
anyonic statistics in strongly interacting systems through both the 
energy spectra and the particle ordering contained in the wave 
functions. As a concrete example we consider using a strongly interacting
impurity in a tunneling experiment to infer the statistical properties
of the majority particles.

\section{Model}
The model Hamiltonian for our $N$-body system has the form
\begin{align}\label{hamil}
H=\sum_{i}\left[-\frac{\hbar^2}{2m}\frac{\partial^2}{\partial x_{i}^{2}}+V(x_i)\right]+g\sum_{i<j}\delta(x_i-x_j),
\end{align}
where $m$ is the mass, $V(x)$ is the external trap potential, and $g$ is 
the interaction strength. Here we assume that all particles have the 
same mass and the same interaction strength which is always parametrized
by $g$. The trap potential length scale is $L$ (box size or harmonic trap length)
which is our basic unit throughout. From $L$ we obtain $\hbar^2/mL^2$
as our unit of energy and likewise we will measure $g$ in units of $\hbar^2/mL$.
The anyonic exchange symmetry implies that \cite{girar2006}
\begin{align}\label{exchange}
\Psi(x_{k},x_{k+1})=-e^{-i\theta\epsilon(x_{k+1}-x_{k})}\Psi(x_{k+1},x_{k}),
\end{align}
where we suppress the dependence on all $N$ coordinates for simplicity 
and $x_{k}$ and $x_{k+1}$ are two adjacent identical (anyonic) particles that we exchange
and $\epsilon(x)=-\epsilon(-x)=1$ ($\epsilon(0)=0$). 
For $\theta=0$ they
are fermions (F) and for $\theta=\pi$ they are bosons (B). The boundary 
conditions are dictated by $V(x)$. It is open boundary conditions, i.e. 
$\Psi$ vanishes at the end of the box for the hard-wall case, while for
the harmonic trap one has gaussian decay at large distance. Periodic 
boundary conditions are not discussed here.

When we discuss two-component
mixtures below it is important to note that there are no symmetry requirements 
between different components, i.e. the wave function may acquire an 
arbitrary phase under exchange of an $A$ and a $B$ particle. 
In Ref.~\cite{santos2012}, solutions with 
symmetric (bosonic) exchange of $A$ and $B$ particles have been discussed. 
We obtain the eigenstates of the Hamiltonian without restrictions
on the exchange of $A$ and $B$. 
These eigenfunctions can have different phases
under exchange of $A$ and $B$, but they are nevertheless eigenstates and
thus the physically relevant states.

A powerful 
feature of our approach to strongly interacting systems \cite{vol2013}, 
is that we obtain these states without using the representation theory of 
symmetry algebras.
As discussed in Refs.~\cite{batchelor2006,batchelor2006b} using the 
Bethe ansatz, the ground state
energy depends $g$ and $\theta$. 
This is also the case here as illustrated in Fig.~\ref{slopes}
for the FF, BF, and BB limits. Our formalism goes beyond the 
Bethe ansatz since it can treat arbitrary external traps.
Introducing several strengths for intra- and interspecies
interactions is an interesting question that has led to recent surprises
\cite{zinner2013} but will not be pursued here. Furthermore, one could include
also odd-parity interactions \cite{girar2004} but we assume that these
are negligible compared to the even-parity ones in Eq.~\eqref{hamil}.

To find the spectrum and the eigenstates for $1/g\to 0$, we construct
a totally antisymmetric wave function, denoted $\Psi_A$, from the $N$ lowest 
single-particle states of the potential $V(x)$. By construction,
$\Psi_A$ vanishes whenever $x_i=x_j$ for any $i,j=1,\ldots,N$.
A general solution of the Schr{\"o}dinger equation for $1/g\to 0$ 
can now be written as 
\begin{align}
\Psi=\sum_n a_n \Psi_A(x_{P_n(1)},\ldots,x_{P_n(N)}),
\end{align}
where the sum runs over all permutations, $P_n$, of the $N$ coordinates.
Solving the Schr{\"o}dinger equation now amounts to finding the 
coefficients $a_n$ which specify the amplitude on each of the orderings
of the $N$ particles. This may be done by noticing that in the 
limit of $1/g\to 0$, the ground state has the largest slope of the energy
as function of $1/g$ (see Fig.~\ref{slopes}a), b) and c)), the first excited the 
second largest slope etc. The slope of the energy may be expressed in 
terms of the $a_n$ coefficients and varied to obtain linear equations 
whose solutions yield the eigenstates \cite{vol2013}. Note that 
the particles are impenetrable in the strict limit where $1/g=0$.
For large but finite $g$ exchange is allowed but suppressed.
In any case the solutions we obtain are accurate to linear order
in $1/g$.

The fact that we are considering identical anyons now impact
the $a_n$ coefficients. To illustrate this, we
consider two adjacent particles with coordinates $x_1$ and $x_2$ 
and assume that $a_1\Psi_A$ is the wave function for $x_1>x_2$ while
$a_2\Psi_A$ is the one for $x_1<x_2$. The contribution to the slope of 
the energy in the limit $1/g\to 0$ is proportional to $|a_1-a_2|^2$ (see
Appendix~\ref{appa} for technical details).
Assuming two identical anyons that 
obey Eq.~\eqref{exchange}, we have $a_1=a_2 e^{i\theta}$ (the minus
sign in Eq.~\eqref{exchange} is due to the antisymmetry of $\Psi_A$).
The contribution becomes $|a_1|^2 4\sin^2(\theta/2)$. Thus the 
slope of the energy and the equations for the eigenfunctions
will now depend on $\theta$. Had we instead considered a pair of non-identical 
$A$ and $B$ particles, then there is no a priory exchange symmetry 
that relates $a_1$ and $a_2$. In that case, the eigenstates of
the Hamiltonian in Eq.~\eqref{hamil} decide what $a_1$ and $a_2$ is.
We note that for $\theta=\pi$ we recover the hard-core boson solutions
of Girardeau \cite{girardeau1960}, while for $\theta=0$ we have identical 
(spinless) fermions. The illustrative example of $N=3$ is discussed in 
Appendix~\ref{appa} below and we refer the reader to that
discussion for the full details.

\begin{figure}[ht!]
\centering
\includegraphics[scale=0.45]{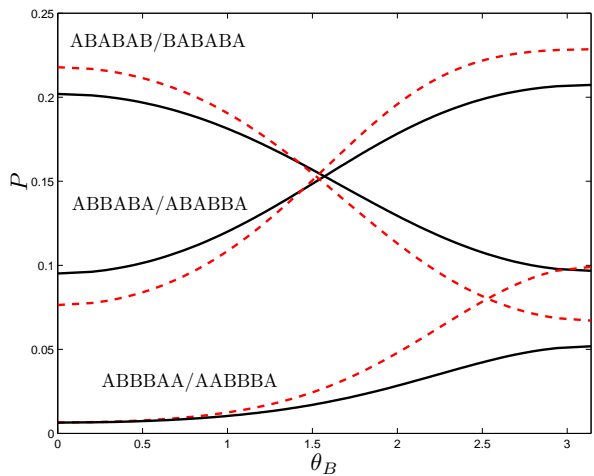}
\caption{(Color online) Same as in Fig.~\ref{slopes}d) but for a balanced 
six-body case where the $A$ particles are fermions ($\theta_A=\pi$)
assuming an external confinement that is a box (solid lines) or a
harmonic trap (dashed lines). The configurations as indicated above
each set of lines. For simplicity we show only the configurations with 
the largest probabilities.}
\label{6balance}
\end{figure}

\section{Balanced systems}
In Fig.~\ref{slopes} we show the $N=4$ case with two 
$A$ and two $B$ particles. Panels a), b), and c)
show the slopes of the energy around $1/g=0$ as the 
statistics changes from Fermi-Fermi (FF) a), across
Bose-Fermi (BF) b), and onto the Bose-Bose (BB) mixture case
in c). Notice the totally antisymmetric state in a) 
which is the horizontal line and corresponds to $a_k=1$
for all orderings. It is only a solution in the FF case.
The slopes have a distinct evolution with statistics
which could in principle be observed by energy measurements
in strongly interacting systems. Fig.~\ref{slopes} assumes
a box trap but only minute quantitative changes occurs if 
one uses harmonic confinement. There has been a lot 
of recent interest in strongly interacting Fermi-Fermi 
\cite{harshman2012,brouzos2013,gharashi2013,sowinski2013,vol2013,cui2014,lindgren2014,deuret2014}
and Bose-Bose mixtures \cite{zollner2008,hao2009,zinner2013,garcia2013}, 
and extended focus on the spatial configuration that such systems 
display for strong interactions. In Fig.~\ref{slopes}d) we 
present the exact results for the ground state configurations
in the limit $1/g\to 0$. For the FF mixture we see a dominant 
antiferromagnetic $ABAB/BABA$ configuration, while the 
BF case has $ABBA$ as the most probable. Finally as we
go to the BB limit, the state originally proposed by
Girardeau \cite{girardeau1960} becomes the exact ground 
state.

The generalized Girardeau type state
proposed in Ref.~\onlinecite{girar2007} has a completely mixed density profile (identical 
to perfect fermionization of four particles) and has been 
shown to agree rather well with a wave function inspired
by a combination of the Bethe ansatz for homogeneous space
and the single-particle solutions of the particular 
trap \cite{fang2011}. This is a kind of hybrid solution 
of the trapped problem. Using our exact solutions 
in the strongly interacting regime one may easily
check that the generalized Girardeau state is a linear 
combination of the eigenstates with slopes shown in 
Fig~\ref{slopes}a (with coefficients that depend on the 
geometry of the trap). It is therefore not connected to 
eigenstates for large but finite interaction strengths
and thus of little experimental relevance.
The numerical Density Matrix Renormalization Group (DMRG) results in Ref.~\onlinecite{fang2011}
seems to agree with a mixed state for very large interaction
strengths which hints at an underlying issue with 
applying DMRG to strongly repulsive particles. It is 
intrinsically variational and will therefore have great
difficulties with the (quasi)-degenerate many-body spectrum
for strong interactions unless one uses exact solutions 
as a guide here. However, we notice that for  
large but not extreme values of the repulsive coupling 
strength Ref.~\onlinecite{fang2011} does indeed find the demixed
ground state that is perfectly consistent with the 
exact result presented here. 

\begin{figure}[ht!]
\centering
\includegraphics[scale=0.45]{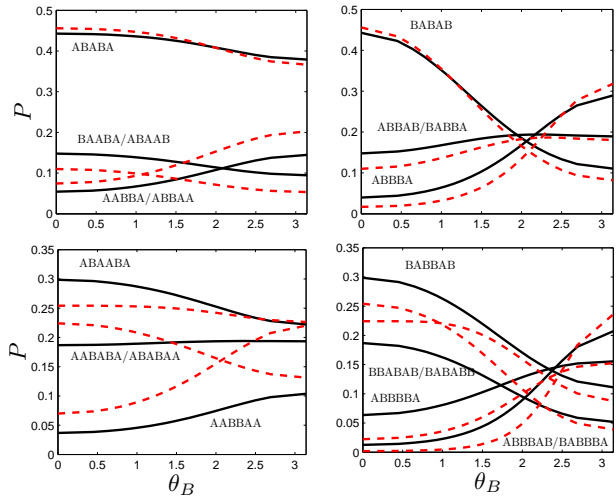}
\caption{(Color online) Imbalanced five- (upper row) and six-body (low row) systems. All 
$A$ particles are fermions ($\theta_A=0$). The configurations follow the lines
in the $\theta_B=0$ limit (left-hand side) from top to bottom. Solid lines are 
for box and dashed lines for harmonic trapping. As in Fig.~\ref{6balance} 
we show only the dominant configurations.}
\label{56full}
\end{figure}

For larger systems the story is similar as we show in 
Fig.~\ref{6balance} with the antiferromagnetic 
dominance being taken over by mixed configurations 
as one goes from FF to BF limits. Note that we only show
the configurations carrying the largest part of the total 
probability.
We omit the results 
as $\theta_A\to\pi$ (BB limit) as they are similar to the 
four-body case in Fig.~\ref{slopes}d).
However, in Fig.~\ref{6balance}
we show results for both a box and a harmonic trap which 
indeed demonstrates that the trap can have decisive influence
on system configuration for $1/g\to 0$ both quantitatively
and qualitatively. In particular, mixed configurations 
dominate the antiferromagnet in the BF limit for harmonic 
but not for box traps. This shows how trap engineering can become
state engineering as first discussed in Ref.~\onlinecite{vol2013}.

\section{Imbalanced systems}
We now explore the interplay of statistics and 
imbalanced in our strongly interacting mixed systems. 
The two upper panels in Fig.~\ref{56full} show 
the cases with three (left) and two (right) fermions ($A$)
mixed with anyons ($B$). Again the FF limit is clearly 
antiferromagnetic, while in the BF limit it depends on 
which particle is in majority. With three $B$ particles, 
the BF system is dominated by the (phase separated) $ABBBA$
configuration, while with only two $B$ particles the system
remains mainly antiferromagnetically ordered ($ABABA$). The 
differences due to the box or harmonic trap are merely quantitative 
in this case. In contrast, for the six-body systems in the two
lower panels of Fig.~\ref{56full}, we do see some qualitative 
changes with external trap, where a harmonic trap enhances
the configuration $AABBAA$ in the BF limit (lower left panel).
Similarly for the lower right panel, we see enhancement of the 
phase separated $ABBBBA$ and $ABBBAB/BABBBA$ configurations, 
and again some qualitative dependence on the trap. We
conclude that the tendency for phase separation for larger 
systems in the BF limit discussed in the introduction seems
to be there but that we identify a crucial dependence on the 
confinement which makes the local density approximation questionable
for smaller systems. An outstanding problem is to extrapolate
the results obtained here to larger system sizes and match the 
few- and many-body limits.  

\begin{figure}[t!]
\centering
\includegraphics[scale=0.43]{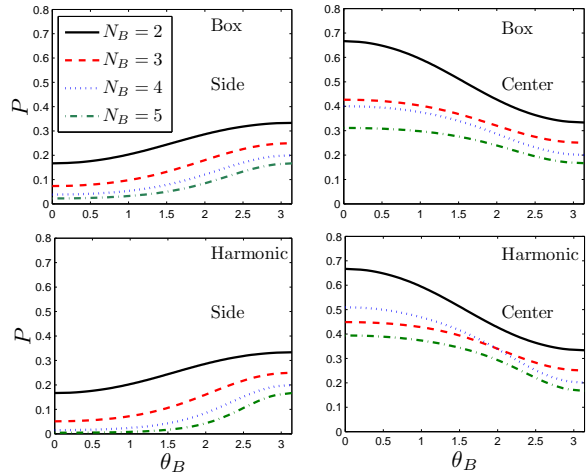}
\caption{(Color online) Probabilities to find an impurity on the 
side (left column) or in the middle (right column) of an 
anyonic system with up to five anyons as function of statistical
parameter, $\theta_B$. The upper row is for a box and the lower row
for a harmonic trap. The middle is defined as either the single 
(for even $N_B$) or the two equivalent central positions (for odd $N_B$).
For $\theta_B=\pi$ all configurations have equal 
probability (BB regime).}
\label{impurity}
\end{figure}

\section{Probing statistics with an impurity}
Finally, we address the case of a single impurity that is strongly
interacting with a number of anyons, $N_B$. As 
discussed above the statistic of the anyons will in general 
influence the energy spectrum and the configurations in the 
system. Measuring the 'fan' of states shown in Fig.~\ref{slopes}
could therefore provide insights into the statistics by comparing
to the theoretical prediction given the trap shape and the 
number of particles. The dependence of the slopes on statistics
has also been identified within the Bethe ansatz approach for 
single-component anyons \cite{batchelor2006b}.

A different approach which can access more information about the system 
is to use tunneling experiments as done recently for an FF mixture \cite{zurn2012}.
In the limit $1/g\to 0$ where the particles become impenetrable, one
can use a simple picture when opening the trap by lowering the trap 
on one side \cite{zurn2012}. Here we may assume that only the particle
located immediately next to the lowered barrier can tunnel. The probability
that this is the impurity can then be approximated by the configurational 
probabilities that we have discussed above. In the left panels of 
Fig.~\ref{impurity} we show this probability for different $\theta_B$
in a box (upper) or harmonic (lower) trap. We see clear variation 
with $\theta_B$ and with $N_B$ which implies that this could be 
used to detect the statistics of 1D anyonic systems. While the 
precise way in which the trap is lowered to allow for tunneling is
except to have a minor quantitative effect, we do not except qualitative
differences. Alternatively, it may be possible to use single site/single
atom resolution quantum gas miscroscopy \cite{bakr2009,sherson2010}
to probe the 1D system locally \cite{simon2011,cheneau2012,fukuhara2013}. 
Here one can probe the probability of 
finding the impurity in the center of the trap which is also very
sensitive to statistics as shown on the right-hand panels in Fig.~\ref{impurity}.
While the experiments cited here have an optical lattice on top of 
the external confinement, this will not qualitatively change our 
predictions. It may change the geometric factors from the confinement
which can be computed using the formulas presented in Ref.~\onlinecite{vol2013}.

\begin{acknowledgments}
I am grateful to K. Sun, O.~I. P\^{a}tu, and A. del Campo for feedback and 
discussions about anyonic systems.
I would like to thank my collaborators A.~G. Volosniev, D.~V. Fedorov, 
A.~S. Jensen, and M. Valiente for all their help and continued work 
on developing our understanding of strongly interacting low-dimensional 
systems. This work was funded by the 
Danish Council for Independent Research DFF Natural Sciences and the 
DFF Sapere Aude program.
\end{acknowledgments}

\appendix 

\section{Illustration of the solution technique}\label{appa}
\begin{figure}[t!]
\centering
\includegraphics[scale=0.33]{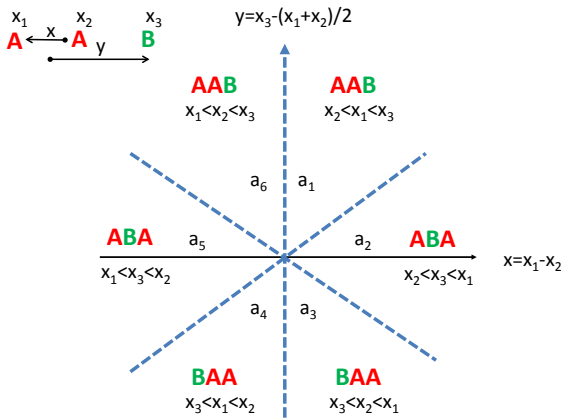}
\caption{(Color online) Schematic of the coordinate space for the three-body problem where 
the $A$ particles are anyons with statistical parameter $\theta$ while 
single the $B$ particle can be considered an impurity.}
\label{3body}
\end{figure}

We now go through the simple example of three particles where two 
are anyons in order to illustrate the differences that arise from 
generalized statistics. We will be very brief and refer the interested
reader to seek further details in Ref.~\onlinecite{vol2013}.
To solve the problem in the limit where $1/g\to 0$ we start from the 
totally antisymmtric wave function, $\Psi_A$, which is zero whenever
any of the three particles overlap in space. We work exclusively 
with the ground state but the technique applies to excited states
as well. The coordinate space
for the three particles is illustrated in Fig.~\ref{3body} where
$A$ are anyons with statistical parameter $\theta$, while $B$ is of 
a different kind (an impurity). 

The most general wave function 
with the correct boundary condition 
is found by taking $\Psi_A$ with a different coefficient,
$a_i$, in the six regions in Fig.~\ref{3body}. This basis is 
complete and we may expand the solutions in the limit $1/g\to 0$ 
in this basis. As shown in Fig.~1 of the main text all solutions are
degenerate in energy when $1/g=0$. We now use the fact that 
as $1/g\to 0$, the ground state (for $g>0$) has the maximum 
slope of the energy as a function of $g$. Using linear perturbation 
theory in $1/g$ or the Hellmann-Feynman theorem, we have $E=E_0-K/g$ where
\begin{align}\label{de}
K=\lim_{g\to\infty}\,g^2\frac{\sum_{i>j}\int  |\Psi|^2\delta(x_i-x_j)\prod_{k=1}^{3}dx_k}{\langle \Psi|\Psi\rangle}.
\end{align}
The full wave function, $\Psi$, is a function of $x_1,x_2,x_3$ 
and consists of the six pieces $a_i\Psi_F$ in Fig.~\ref{3body}. 
The sum, $\sum_{i>j}$, runs over all pairs according to the Hamiltonian
in Eq.~(1) of the main text. $\langle \Psi|\Psi\rangle$ is the normalization integral. 
We now eliminate $g$ by using the zero-range boundary condition
\begin{align}\label{bound}
-\frac{1}{2g}\left[ \frac{\partial\Psi}{\partial x_+}-\frac{\partial\Psi}{\partial x_-}\right]=\Psi,
\end{align}
where $x_{\pm}=\pm(x_i-x_j)$.

After some calculations along the lines described in Ref.~\onlinecite{vol2013}, 
the expression for $K$ becomes 
\begin{align}\label{K21}
K=K_0\frac{(a_1-a_2)^2+(a_2-a_3)^2+(a_{1}^{2}+a_{3}^{2})|1-e^{-i\theta}|}{a_{1}^{2}+a_{2}^{2}+a_{3}^{2}},
\end{align}
where $K_0$ is a geometric factor that depends on the trapping potential and
the coefficients, $a_i$, are real numbers.
Notice that for larger systems there are more than one of these factors in the 
result \cite{vol2013}, but for three-body systems it can be taken 
outside for the parity invariant box and harmonic potentials we work with 
here. By variation of $K$ with respect to $a_1$, $a_2$, and $a_3$, one 
can obtain the eigenstates that are adiabatically connected to the eigenstate
for large but finite $g$ as well as the slope of the energy to linear order 
in $1/g$.

As discussed, the decisive quantity that determines the wave functions that 
are adiabatically connected eigenstates in the limit $1/g=0$
is the slope of the energy, $K$. If we describe the anyons as
strictly hard core particles this means they are to be regarded
as ideal fermions and will make no contribution to $K$. 
This reduces the problem to that of 
two identical fermions and an impurity, and this is true no 
matter what value the anyonic exchange parameter, $\theta$, takes.
In turn one would not be able to recover the correct Bose-Fermi mixture
limit discussed in detail for the four-body system in the main text. This 
implies that one needs to consider the anyons when calculating
the energies even in the 'hard core' limit $1/g\to 0$ in order to 
have a model that matches the behavior in the known limiting 
cases where $\theta=0$ or $\theta=\pi$. Our approach is therefore
closely related to the work using the Bethe ansatz in 
Refs.~\onlinecite{batchelor2006,batchelor2006b}, and we also obtain 
a strong coupling expansion of the energy which depends on 
$\theta$ (see Eq.~(10) of Ref.~\cite{batchelor2006b}). However,
our formalism goes beyond the Bethe ansatz in being able to 
handle arbitrary confining geometries without resorting to 
the local density approximation.

\end{document}